\documentclass[12pt]{article}
\usepackage{newtxtext,newtxmath}
\usepackage{graphicx}

\usepackage[letterpaper,margin=1in]{geometry}

\renewenvironment{abstract}
	{\quotation}
	{\endquotation}

\date{}

\makeatletter
\renewcommand{\fnum@figure}{\textbf{Figure \thefigure}}
\renewcommand{\fnum@table}{\textbf{Table \thetable}}
\makeatother

\usepackage{scicite}

\usepackage{url}

\def\scititle{
	High-speed phase-encoded quantum secure direct communication over 11.4 km heterogeneous free-space and fiber links
}
\title{\bfseries \boldmath \scititle}

\author{
	Ze-Zhou Sun$^{1,2,\ast}$, Yuan-Bin Cheng$^{1,2,\ast}$, Yu-Chen Liu$^{1,2,\ast}$, Jianxing Guo$^{1}$,\\ Xiao-Tian Song$^{1}$, Wei Zhang$^{1}$, Dong Pan$^{1,\dagger}$ and Gui-Lu long$^{1,2,3,4,\dagger}$\and
	\small$^{1}$Beijing Academy of Quantum Information Sciences, Beijing 100193, China\and
	\small$^{2}$State Key Laboratory of Low-dimensional Quantum Physics and Department of Physics,\\ \small Tsinghua University, Beijing 100084, China\and
		\small$^{3}$Frontier Science Center for Quantum Information, Beijing 100084, China\and
			\small$^{4}$Beijing National Research Center for Information Science and Technology, Beijing 100084, China\and
	\small$^\dagger$Corresponding authors' email: pandong@baqis.ac.cn; gllong@mail.tsinghua.edu.cn\and
	\small$^\ast$These authors contributed equally to this work
}

\begin{document} 

\maketitle

\begin{abstract} \bfseries \boldmath
Robust quantum transmission is driving a new paradigm in space–ground quantum networking. Although phase encoding has been widely adopted in terrestrial fiber channels, it has long been considered unsuitable for free-space quantum communication. Here, we demonstrate phase-encoded quantum communication over 1400 m of urban free space. The system maintained stable operation for nearly one hour, achieving 99.07\% interference visibility and an average quantum bit error rate of 2.38\%. The free-space quantum states were directly coupled into the fiber and transmitted over an additional 10 km, confirming seamless interoperability across different media. We further show that turbulence-induced phase drifts between successive picosecond pulses can be effectively compensated. A cascaded-link model and numerical simulations indicate feasibility over free-space distances exceeding 30 km, underscoring the potential for satellite-to-ground quantum links. This work establishes the viability of phase encoding in free-space quantum networks, simplifying cross-medium integration and enabling compatibility with existing classical infrastructures.
\end{abstract}

\noindent
A variety of quantum encoding schemes—including polarization, phase, time-bin, and quadrature of light—have been proposed and successfully demonstrated for quantum communication~\cite{lloyd2004infrastructure,kimble2008quantum,wehner2018quantum,long2022evolutionary,pan2024evolution}. Among them, polarization encoding has been most widely employed in free-space links due to its simplicity and the intrinsic stability of polarization states in atmospheric channels~\cite{buttler1998free,schmitt2007experimental,vallone2015experimental,zhang2017quantum,avesani2021full}. However, polarization mode dispersion, birefringence, and depolarization fundamentally limit its performance in optical fibers. By contrast, time-bin and phase encoding are widely used in single-mode fiber networks, where they offer high stability, yet phase encoding is often deemed unsuitable for free-space transmission. If phase encoding were feasible in free space, it could greatly simplify infrastructure and reduce costs.

However, phase encoding in free-space channels faces three major challenges. First, the polarization of photonic states may drift, necessitating both active and passive compensation, while the complexity of atmospheric channels further aggravates phase instability. Second, turbulence-induced spatial perturbations disrupt temporal-mode measurements~\cite{jin2018demonstration}. Third, link-loss fluctuations degrade interference visibility~\cite{castillo2023multiprotocol}. As quantum communication advances toward satellite-to-ground links, achieving high interference visibility remains a critical challenge~\cite{vallone2016interference}. 

The transmission loss, decoherence, and noise mechanisms of quantum states vary markedly across physical media, making cross-medium interconnection a persistent challenge. While free-space and fiber channels each offer distinct advantages, existing quantum communication systems remain dominated by point-to-point links tailored to specific encodings. The absence of architectures that seamlessly span different media has hindered large-scale quantum internet deployment and created an urgent demand for universal solutions.
The inherently low energy of quantum states, the limited performance of current devices, and the short transmission windows of space-based links~\cite{bedington2017progress} place efficient use of quantum resources at the center of research. Recent strategies emphasize resource reuse through multifunctional integration, such as combining quantum key distribution (QKD) with quantum time synchronization~\cite{xiang2025towards} and exploiting the same photon states for quasi quantum secure direct communication (QSDC) and QKD~\cite{pan2025simultaneous}. These approaches open new pathways to enhance resource efficiency at the protocol layer of quantum networks~\cite{cao2022evolution,illiano2022quantum,suncheng25}.

Here we present a hybrid free-space–fiber quantum communication system based on a 1.25-GHz weak coherent light source with pure phase modulation. The system enables nearly one hour of continuous and stable transmission over 1.4 km of urban free space and 10 km of optical fiber. High interference visibility and low quantum bit error rates are maintained for four-phase-encoded states, defying the conventional expectation that atmospheric turbulence severely disrupts phase stability. A one-way quasi-QSDC~\cite{pan2025simultaneous} is achieved with a communication rate of 4.22 kbps. The encoding’s compatibility with fiber backends reduces network heterogeneity and system complexity, while the patch-cord optical interface extends the trusted boundary from telescope to internal devices, enhancing both security and deployment flexibility. Finally, we establish a qubit transmission model for this cross-media architecture, providing a theoretical foundation for future space–ground quantum networks.

\section*{Methods}
\textbf{The running protocol.}
We implement the simultaneous transmission of information and key exchange (STIKE) protocol~\cite{pan2025simultaneous} using four-phase encoding with Faraday–Sagnac–Michelson interferometers. The protocol comprises three main steps:
(1)	Encoding and preparation. To transmit quantum information between two distant parties, conventionally Alice and Bob, Alice first preprocesses the message using forward error correction, spread-spectrum coding, one-time-pad encryption, and masking. The resulting codewords are then encoded into quantum states, and photons are sent to Bob.
(2)	Measurement and security check. Upon receiving the photon sequence, Bob randomly selects measurement bases (Z or X) for each photon and publicly announces the corresponding timestamps and basis choices. A subset of basis-matched events is then jointly sampled to calculate the quantum bit error rate (QBER). If the QBER is below a predetermined threshold, the transmission is considered secure, and the protocol proceeds~\cite{WANG202191}.
(3) Decoding and post-processing. Both parties retain only the events measured in the same basis. Bob then performs three operations: (i) sequential unmasking, decryption, dispreading, and error-correction decoding to recover Alice’s message, realizing quasi-QSDC; (ii) extraction of raw keys from the retained photons and Alice’s codewords for QKD, generating fresh encryption keys for subsequent quasi-QSDC sessions; (iii) recovery of keys associated with lost or discarded photons, enabled by channel monitoring and masking, allowing full recycling of consumed keys.

Relevant performance metrics, including the QBER, secure communication rate $C_s$, key consumption rate, key generation rate, and the percentage of key recycling $P_{\rm rec}$, are recorded in real time during transmission by dedicated software. The key consumption rate is the rate at which keys are consumed for the retained photons. The percentage of key recycling refers to the fraction of keys recovered relative to the total number consumed in the encrypted codewords, and is theoretically given by $P_{\rm rec} = 1 - Q_\mu/2$ in the absence of eavesdropping, where $Q_\mu$ denotes the gain of the signal states.

\begin{figure}[h]
\centerline{\includegraphics[width=\linewidth]{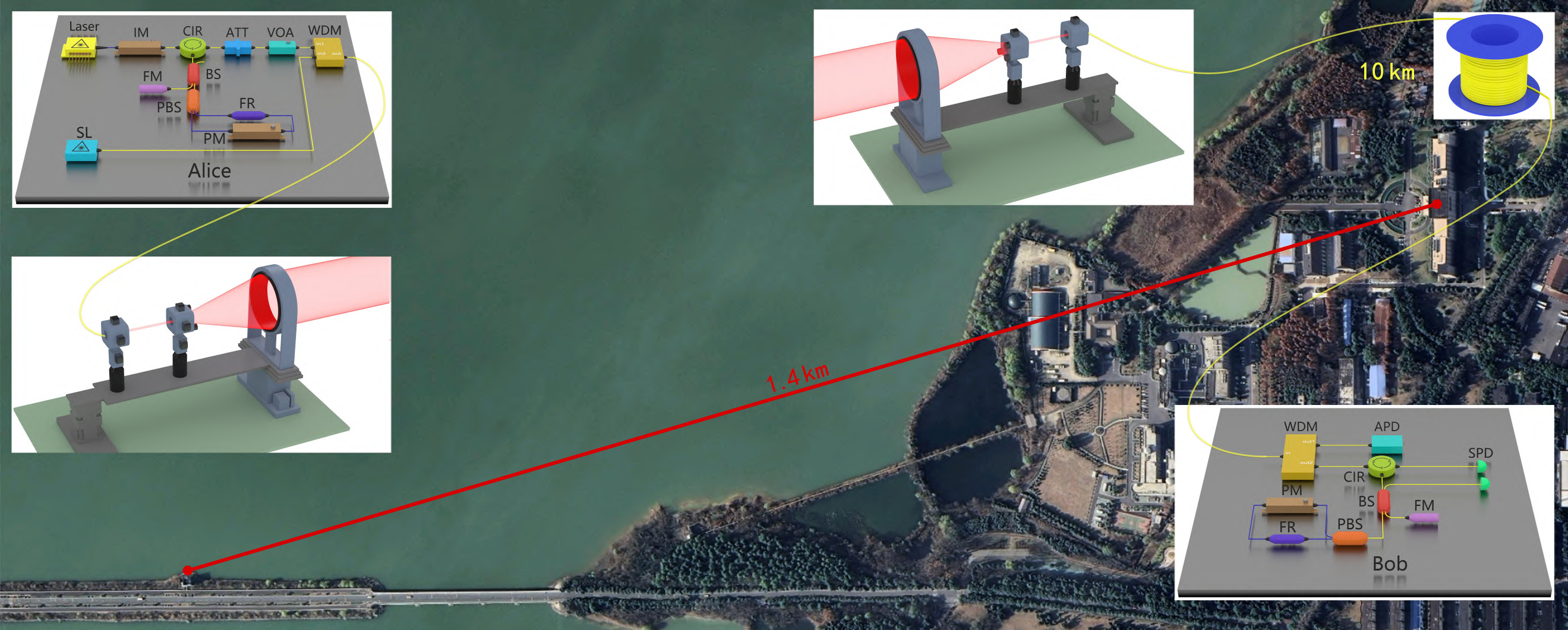}}
\caption{\label{Fig_place} Schematic of our experimental setup featuring a 1.4 km free-space atmospheric channel over a lake surface combined with a 10 km optical fiber link in Hefei, China. IM, intensity modulator; CIR, circulator; BS, beam splitter; FM, Faraday mirror; PBS, polarization beam splitter; FR, Faraday rotator; PM, phase modulator; ATT, attenuator; VOA, variable optical attenuator; SPD, single photon detector; WDM, wavelength division multiplexer; SL, synchronization laser; APD, avalanche photon diode. }
\end{figure}

\noindent\textbf{System details.} Fig.~\ref{Fig_place} shows the experimental setup and environment. A 1.4 km free-space channel was established across an urban lake in Hefei, China, between the transmitter (Alice) at ($31^\circ 53^\prime 36.4^{\prime\prime}\mathrm{N},117^\circ 09^\prime 54.7^{\prime\prime}\mathrm{E}$) and the receiver (Bob) at ($31^\circ 54^\prime 18.6^{\prime\prime}\mathrm{N},117^\circ 09^\prime 38.5^{\prime\prime}\mathrm{E}$). Photons arriving at Bob were coupled into a 10 km optical fiber using a passive telescope system with a 120 mm entrance pupil, 27.1$\times$ magnification, a total focal length of 395 mm, and an optical path length of 330 mm. A triplet fiber-optic collimator was employed for efficient free-space–fiber coupling. Experiments were performed under two representative atmospheric conditions: (i) evening of May 26, 2025, wind level 2, $23^\circ\mathrm{C}$, 60\% humidity, visibility 24.1 km; and (ii) early morning of May 28, 2025, wind level 1, $29^\circ\mathrm{C}$, 40\% humidity, visibility 14 km.

Alice employs a 1.25 GHz laser with a pulse width of 50 ps and a central wavelength of 1549.32 nm to generate weak coherent pulses. An intensity modulator (IM), driven by a field-programmable gate array, modulates the laser output by applying pulse voltages of varying amplitudes, producing signal states ($\mu = 0.71$), decoy states ($\nu_1 = 0.28$), and vacuum states ($\nu_2 = 0$) in a ratio of 30:2:1. The IM is connected to a Faraday–Sagnac–Michelson interferometer with unequal arms via a circulator to perform encoding at the transmitter. Two cascaded phase modulators (PMs) randomly modulate phases of $\{$0 or $\pi/2$$\}$ and $\{$0 or $\pi$$\}$ respectively according to the transmitted information, thereby enabling random four-phase modulation through their combination. After passing through the transmitting Faraday–Sagnac–Michelson interferometer loop, each pulse is split into two pulses separated by 400 ps and wavelength-division multiplexed with a 500 kHz clock synchronization signal, which has a central wavelength of 1550.92 nm. Alice and Bob perform authenticated classical communication via a wireless Wi-Fi bridge. Bob employs a dual-channel InGaAs/InP single-photon detector (SPD) to detect photons. The detector operates at 1.25 GHz with a detection efficiency of 20\%, a dark count rate of $1 \times 10^{-6}$, and an afterpulse probability of 1\%. The inherent optical loss at Bob’s receiving end is 6.5 dB. The system adopts a frame-by-frame transmission scheme to facilitate reliable communication. Each frame consists of 125 bytes and is transmitted with a spreading ratio of 1:1920, enhancing transmission robustness under varying channel loss conditions.

\section*{Results}
\begin{figure}[htbp]
\centerline{\includegraphics[width=\linewidth]{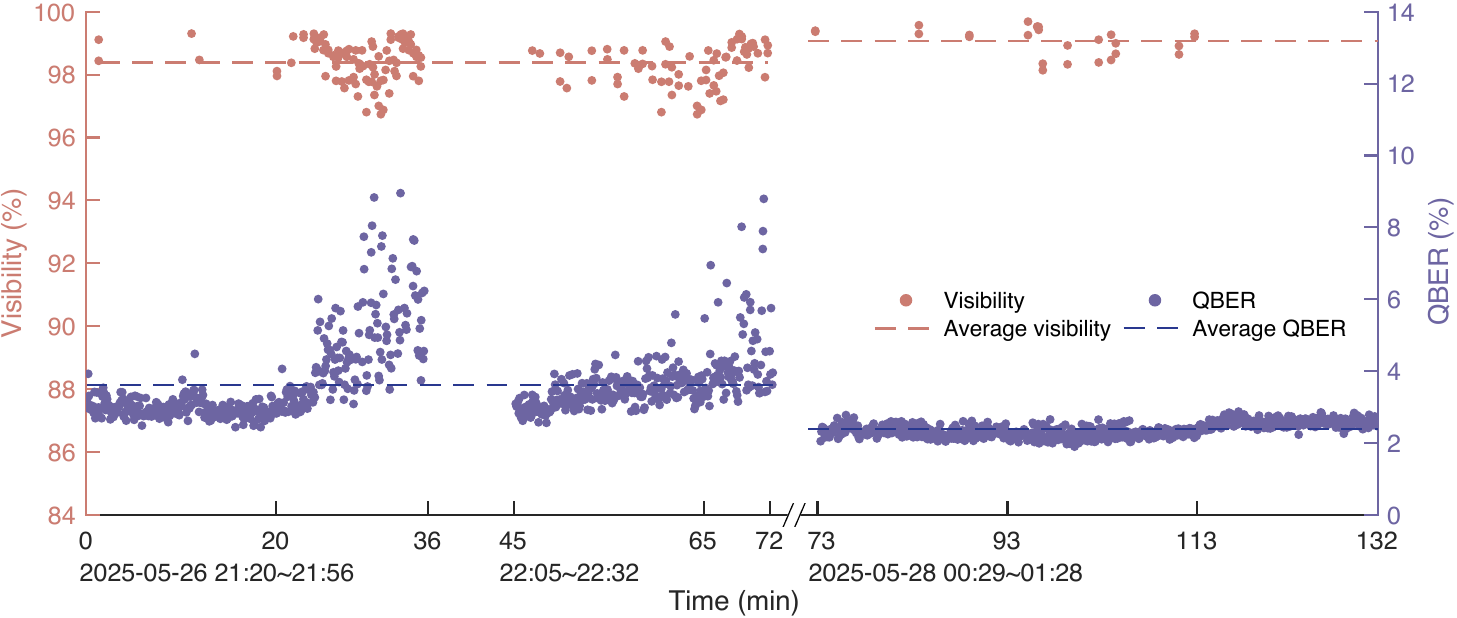}}
\caption{\label{Fig_inte} 
Interference visibility and QBER as functions of time. The average QBERs over the two experimental days were 3.61\% and 2.38\%, respectively. On the first day of operation the beam deviated from alignment at the 36th minute causing a link disconnection. The optical path was successfully recalibrated, which restored communication at the 45th minute.}
\end{figure}

In Fig.~\ref{Fig_inte}, we report the measured interference visibility and QBER during transmission, with the corresponding channel loss provided in Extended Data Fig.~\ref{Fig_loss}. The average visibilities over the two-day experiment reached 98.38\% and 99.07\%, respectively, comparable to state-of-the-art fiber-based quantum communication~\cite{korzh2015provably,wang2018practical}. These results confirm the feasibility of phase-encoded quantum transmission over both optical fiber and free-space channels. Notably, the interference visibility significantly surpasses that of previous free-space demonstrations~\cite{pan2020experimental,sajeed2021observing,cocchi2025time}, despite the increased number of transmitted quantum states and the longer free-space distance (Extended Data Fig.~\ref{Fig_INT}). During intervals when QBER temporarily exceeded 5\%, interference visibilities were actively scanned to verify device integrity, meaning that the reported values represent a conservative lower bound. On the first day, the free-space link exhibited pronounced instability, with extreme QBER fluctuations and eventual transmission failure caused by the absence of an acquisition, pointing, and tracking (APT) system. By contrast, the second day demonstrated uninterrupted long-distance operation, where our built-in phase compensation mechanism effectively restored robustness against environmental turbulence.

Hence, we demonstrate that 1.25-GHz phase-encoded quantum states can withstand turbulence-induced phase drifts through active compensation. Detector counts yield the interference visibility, from which the actual phase difference is obtained. Comparing this with the ideal value provides the voltage offset associated with phase drift, which is subsequently applied to the ideal phase setting to realize dynamic feedback compensation. This robustness arises because the typical timescale of turbulence-induced fluctuations (0.01–0.1 s~\cite{gisin2002quantum}) is orders of magnitude longer than the 400 ps separation between adjacent pulses. Consequently, consecutive pulses encoding the relative phase experience nearly identical atmospheric conditions, and the slowly varying turbulence has only a negligible impact on phase stability. Coupling the free-space signal into single-mode fibers further mitigates spatial perturbations via spatial-mode filtering, which can be enhanced with adaptive optics. Moreover, adopting a one-way protocol~\cite{pan2025simultaneous} reduces overall system loss compared with the two-way scheme~\cite{pan2020experimental}, while independent suppression of channel-loss fluctuations ensures link stability. Together, these measures enable the phase-encoding system to sustain high and stable interference visibility under atmospheric turbulence.

	
\begin{figure}[htbp]
\centerline{\includegraphics[width=\linewidth]{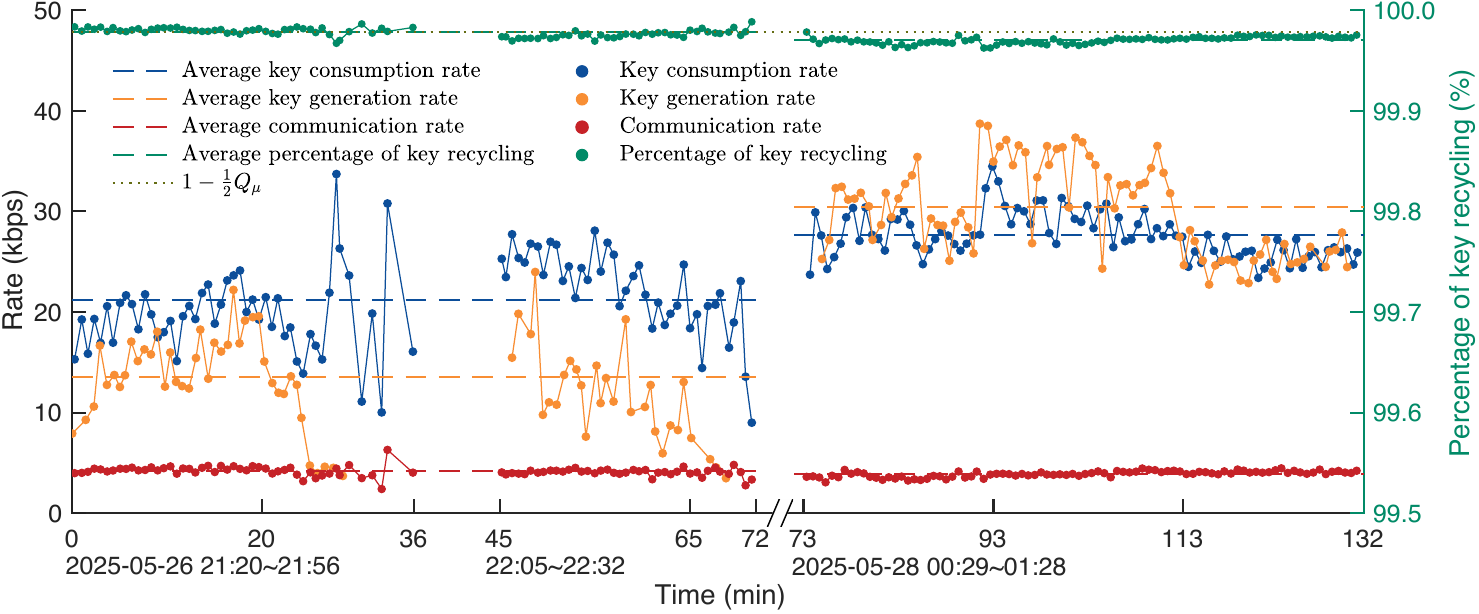}}
\caption{\label{Fig_Rate} 
Transmission performance versus time.
The average communication rates over the two days were 4.22 kbps and 3.90 kbps, the average key generation rates were 13.51 kbps and 30.42 kbps, the average key consumption rates were 21.18 kbps and 27.68 kbps, and the average key recovery percentages were 99.98\% and 99.97\%, respectively.}
\end{figure}

Fig.~\ref{Fig_Rate} illustrates the transmission performance of the STIKE protocol. As the overall channel losses were comparable on the two experimental days (see Extended Data Fig.~\ref{Fig_loss} in the Supplementary Information), the system achieved similar communication rates of 4.22 kbps and 3.90 kbps, with corresponding key consumption rates of 21.18 kbps and 27.68 kbps. A comparison with Figs.~\ref{Fig_inte} and \ref{Fig_Rate} reveals that during intervals of rapidly increasing QBER, the communication rate points become noticeably sparse and the key generation rate gradually declines. This behavior is consistent with the degradation of link quality in the absence of an APT system. Consequently, poor link quality reduces the number of valid communication rate points. Under conditions of strong atmospheric turbulence, the system operated with a key generation rate of 13.51 kbps, which was lower than the consumption rate of 21.18 kbps, leading to a deficit of 7.67 kbps in the key pool. By contrast, during the second day, the channel remained stable with relatively steady QBER, enabling the system to achieve a key generation rate of 30.42 kbps, surpassing the consumption rate of 27.68 kbps and yielding a net surplus of 2.74 kbps for key pool replenishment. Over the two days, the system achieved average key recycling efficiencies of 99.9\% and 99.97\%, respectively, in excellent agreement with the theoretical prediction of $1-Q_\mu/2$.

\begin{figure}[htpb]
\centerline{\includegraphics[width=14cm]{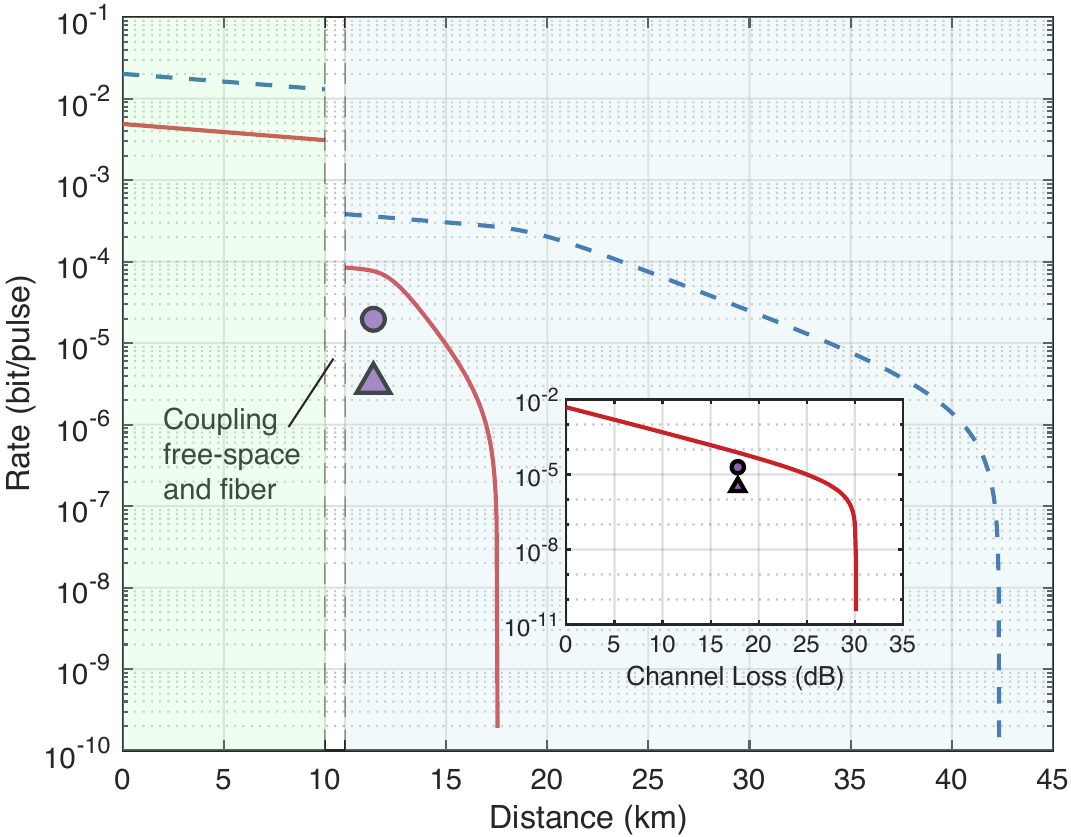}}
\caption{\label{Fig_Capacity}
Simulation (lines) and experimental (symbols) results for the rate. The triangular and circular markers indicate the experimental communication rate and key generation rate, respectively. The solid curve shows the relationship between rate and distance derived from our experimental parameters. The dotted curve is calculated using a detector efficiency of 80\%, a telescope system conversion efficiency of $10^{-15.4/10}$, and a receiving telescope aperture of 1 m.}
\end{figure}

Fig.~\ref{Fig_Capacity} presents the simulated rate bounds alongside our experimental results for the STIKE protocol over a hybrid fiber–free-space channel. Details of the simulation model are provided in the Supplementary Information. The fiber link was set to 10 km, which is representative of practical local-area transmission and can be adjusted as required. The results indicate that the key generation rate of our system is already close to the theoretical limit, whereas the communication rate still falls short of the corresponding bound. To mitigate rapid fluctuations in the free-space channel loss, we employed a large spreading factor, with the resulting coding redundancy inevitably reducing the communication rate. This highlights the trade-off between accommodating channel-loss variations and maintaining communication efficiency in free-space links. As shown in Fig.~\ref{Fig_Rate}, a relatively stable communication rate was nevertheless achieved under the larger spreading factor. Moreover, the simulation results suggest that the proposed cascaded transmission architecture holds strong potential for satellite–ground interactive quantum communication, as indicated by the dashed curves. Free-space transmission over more than 30 km exceeds the effective thickness of the atmosphere.

\section*{Discussion and outlook}
We developed a four-phase-encoded quantum state transmission system based on a Faraday-Sagnac-Michelson interferometer and, for the first time, experimentally demonstrated stable quasi-QSDC over hybrid long-distance free-space and fiber channels, achieving both high interference visibility and low QBER. Our results extend free-space QSDC from meter-scale laboratory demonstrations to the kilometer scale under urban atmospheric conditions, achieving a record transmission rate of 4.22 kbps. This represents an eightfold improvement in transmission rate and a 140-fold extension in distance compared with existing free-space QSDC implementations~\cite{pan2020experimental}. These advances demonstrate the feasibility of QSDC for transmitting text, images, and real-time voice over free-space channels. 

The STIKE protocol operates flexibly in multiple modes according to channel conditions and communication requirements. It can sustain secure communication by allocating no codeword space to random numbers when self-sufficiency in key generation is achieved, or by allocating a portion of the codeword to random numbers when key consumption exceeds key generation, a feature that is particularly important under prolonged adverse link conditions.

The constructed system exhibits notable advantages in resource efficiency, architectural simplicity, and reduced device heterogeneity, while also demonstrating excellent scalability and practical potential for space–ground integrated network deployment. This design enables a single quantum communication device to operate seamlessly across both free space and fiber channels, allowing plug-and-play switching between them and thus realizing cross-medium multiplexing of communication hardware. Furthermore, the phase-encoded architecture is inherently compatible with chip-level integration~\cite{luo2023recent,kwek2021chip}, since on-chip propagation naturally preserves phase information and requires lower power consumption. The preparation, transmission, and interference measurement of qubits based on the phase difference between consecutive pulses form the essential infrastructure for quantum communication~\cite{inoue2002differential,yu2024time,wang2025time}, distributed optical quantum computing~\cite{humphreys2013linear,liu2024nonlocal,rubeling2025fiber}, and quantum sensing~\cite{sajeed2021observing,he2025experimental}. Our results are expected to provide a universal architecture for cross-medium, long-distance quantum information transmission, paving the way toward integrated quantum information processing networks~\cite{fan2025quantum}.

\clearpage 

\bibliography{mybib}

\begin{thebibliography}{10}
\providecommand{\url}[1]{\texttt{#1}}
\expandafter\ifx\csname urlstyle\endcsname\relax
  \providecommand{\doi}[1]{doi:\discretionary{}{}{}#1}\else
  \providecommand{\doi}{doi:\discretionary{}{}{}\begingroup
  \urlstyle{rm}\Url}\fi

\bibitem{lloyd2004infrastructure}
S.~Lloyd, \emph{et~al.}, Infrastructure for the quantum Internet. \emph{ACM
  SIGCOMM Computer Communication Review} \textbf{34}~(5), 9--20 (2004).

\bibitem{kimble2008quantum}
H.~J. Kimble, The quantum internet. \emph{Nature} \textbf{453}~(7198),
  1023--1030 (2008).

\bibitem{wehner2018quantum}
S.~Wehner, D.~Elkouss, R.~Hanson, Quantum internet: a vision for the road
  ahead. \emph{Science} \textbf{362}~(6412), eaam9288 (2018).

\bibitem{long2022evolutionary}
G.-L. Long, \emph{et~al.}, An evolutionary pathway for the quantum internet
  relying on secure classical repeaters. \emph{IEEE Network} \textbf{36}~(3),
  82--88 (2022).

\bibitem{pan2024evolution}
D.~Pan, \emph{et~al.}, The evolution of quantum secure direct communication: on
  the road to the qinternet. \emph{IEEE Communications Surveys \& Tutorials}
  \textbf{26}~(3), 1898--1949 (2024).

\bibitem{buttler1998free}
W.~Buttler, \emph{et~al.}, Free-space quantum-key distribution. \emph{Physical
  Review A} \textbf{57}~(4), 2379 (1998).

\bibitem{schmitt2007experimental}
T.~Schmitt-Manderbach, \emph{et~al.}, Experimental demonstration of free-space
  decoy-state quantum key distribution over 144 km. \emph{Physical Review
  Letters} \textbf{98}~(1), 010504 (2007).

\bibitem{vallone2015experimental}
G.~Vallone, \emph{et~al.}, Experimental satellite quantum communications.
  \emph{Physical Review Letters} \textbf{115}~(4), 040502 (2015).

\bibitem{zhang2017quantum}
W.~Zhang, \emph{et~al.}, Quantum secure direct communication with quantum
  memory. \emph{Physical Review Letters} \textbf{118}~(22), 220501 (2017).

\bibitem{avesani2021full}
M.~Avesani, \emph{et~al.}, Full daylight quantum-key-distribution at 1550 nm
  enabled by integrated silicon photonics. \emph{npj Quantum Information}
  \textbf{7}~(1), 93 (2021).

\bibitem{jin2018demonstration}
J.~Jin, \emph{et~al.}, Demonstration of analyzers for multimode photonic
  time-bin qubits. \emph{Physical Review A} \textbf{97}~(4), 043847 (2018).

\bibitem{castillo2023multiprotocol}
A.~T. Castillo, U.~Zanforlin, G.~Buller, R.~Donaldson, Multiprotocol quantum
  key distribution receiver for free space, in \emph{Quantum Technology:
  Driving Commercialisation of an Enabling Science III} (SPIE), vol. 12335
  (2023), pp. 79--83.

\bibitem{vallone2016interference}
G.~Vallone, \emph{et~al.}, Interference at the single photon level along
  satellite-ground channels. \emph{Physical Review Letters} \textbf{116}~(25),
  253601 (2016).

\bibitem{bedington2017progress}
R.~Bedington, J.~M. Arrazola, A.~Ling, Progress in satellite quantum key
  distribution. \emph{npj Quantum Information} \textbf{3}~(1), 30 (2017).

\bibitem{xiang2025towards}
X.~Xiang, \emph{et~al.}, Towards a Function-Scalable Quantum Network With
  Multiplexed Energy-Time Entanglement. \emph{Laser \& Photonics Reviews} p.
  e00658 (2025).

\bibitem{pan2025simultaneous}
D.~Pan, \emph{et~al.}, Simultaneous transmission of information and key
  exchange using the same photonic quantum states. \emph{Science Advances}
  \textbf{11}~(8), eadt4627 (2025).

\bibitem{cao2022evolution}
Y.~Cao, \emph{et~al.}, The evolution of quantum key distribution networks: On
  the road to the qinternet. \emph{IEEE Communications Surveys \& Tutorials}
  \textbf{24}~(2), 839--894 (2022).

\bibitem{illiano2022quantum}
J.~Illiano, M.~Caleffi, A.~Manzalini, A.~S. Cacciapuoti, Quantum internet
  protocol stack: A comprehensive survey. \emph{Computer Networks}
  \textbf{213}, 109092 (2022).

\bibitem{suncheng25}
Z.-Z. Sun, \emph{et~al.}, Quantum communication network routing with circuit
  and packet switching strategies. \emph{IEEE Journal on Selected Areas in
  Communications} \textbf{43}~(5), 1887--1900 (2025),
  \doi{10.1109/JSAC.2025.3543524}.

\bibitem{WANG202191}
C.~Wang, Quantum secure direct communication: Intersection of communication and
  cryptography. \emph{Fundamental Research} \textbf{1}~(1), 91--92 (2021),
  \doi{https://doi.org/10.1016/j.fmre.2021.01.002}.

\bibitem{korzh2015provably}
B.~Korzh, \emph{et~al.}, Provably secure and practical quantum key distribution
  over 307 km of optical fibre. \emph{Nature Photonics} \textbf{9}~(3),
  163--168 (2015).

\bibitem{wang2018practical}
S.~Wang, \emph{et~al.}, Practical gigahertz quantum key distribution robust
  against channel disturbance. \emph{Optics Letters} \textbf{43}~(9),
  2030--2033 (2018).

\bibitem{pan2020experimental}
D.~Pan, \emph{et~al.}, Experimental free-space quantum secure direct
  communication and its security analysis. \emph{Photonics Research}
  \textbf{8}~(9), 1522--1531 (2020).

\bibitem{sajeed2021observing}
S.~Sajeed, T.~Jennewein, Observing quantum coherence from photons scattered in
  free-space. \emph{Light: Science \& Applications} \textbf{10}~(1), 121
  (2021).

\bibitem{cocchi2025time}
S.~Cocchi, \emph{et~al.}, Time-bin encoding quantum key distribution in
  free-space horizontal links during nighttime and daytime. \emph{Optica
  Quantum} \textbf{3}~(4), 346--350 (2025).

\bibitem{gisin2002quantum}
N.~Gisin, G.~Ribordy, W.~Tittel, H.~Zbinden, Quantum cryptography.
  \emph{Reviews of Modern Physics} \textbf{74}~(1), 145 (2002).

\bibitem{luo2023recent}
W.~Luo, \emph{et~al.}, Recent progress in quantum photonic chips for quantum
  communication and internet. \emph{Light: Science \& Applications}
  \textbf{12}~(1), 175 (2023).

\bibitem{kwek2021chip}
L.-C. Kwek, \emph{et~al.}, Chip-based quantum key distribution. \emph{AAPPS
  Bulletin} \textbf{31}~(1), 15 (2021).

\bibitem{inoue2002differential}
K.~Inoue, E.~Waks, Y.~Yamamoto, Differential phase shift quantum key
  distribution. \emph{Physical Review Letters} \textbf{89}~(3), 037902 (2002).

\bibitem{yu2024time}
H.~Yu, A.~O. Govorov, H.-Z. Song, Z.~Wang, Time-encoded photonic quantum
  states: generation, processing, and applications. \emph{Applied Physics
  Reviews} \textbf{11}~(4) (2024).

\bibitem{wang2025time}
J.~Wang, \emph{et~al.}, Time-bin encoded quantum key distribution over 120 km
  with a telecom quantum dot source. \emph{arXiv preprint arXiv:2506.15520}
  (2025).

\bibitem{humphreys2013linear}
P.~C. Humphreys, \emph{et~al.}, Linear optical quantum computing in a single
  spatial mode. \emph{Physical Review Letters} \textbf{111}~(15), 150501
  (2013).

\bibitem{liu2024nonlocal}
X.~Liu, \emph{et~al.}, Nonlocal photonic quantum gates over 7.0 km.
  \emph{Nature Communications} \textbf{15}~(1), 8529 (2024).

\bibitem{rubeling2025fiber}
P.~R{\"u}beling, R.~Johanning, J.~Heine, O.~V. Marchukov, M.~Kues, Fiber
  transmission of cluster states via multi-level time-bin encoding. \emph{arXiv
  preprint arXiv:2507.01497}  (2025).

\bibitem{he2025experimental}
W.~He, \emph{et~al.}, Experimental secure entanglement-free quantum remote
  sensing over 50 km of optical fiber. \emph{Physical Review A}
  \textbf{111}~(6), 062607 (2025).

\bibitem{fan2025quantum}
Y.-R. Fan, \emph{et~al.}, Quantum entanglement network enabled by a
  state-multiplexing quantum light source. \emph{Light: Science \&
  Applications} \textbf{14}~(1), 189 (2025).

\bibitem{karakosta2025free}
I.~Karakosta-Amarantidou, R.~Yehia, M.~Schiavon, Free-space model for a
  balloon-based quantum network. \emph{Physical Review Research}
  \textbf{7}~(2), 023199 (2025).

\bibitem{bourgoin2015free}
J.-P. Bourgoin, \emph{et~al.}, Free-space quantum key distribution to a moving
  receiver. \emph{Optics Express} \textbf{23}~(26), 33437--33447 (2015).

\bibitem{vasylyev2016atmospheric}
D.~Vasylyev, A.~Semenov, W.~Vogel, Atmospheric quantum channels with weak and
  strong turbulence. \emph{Physical Review Letters} \textbf{117}~(9), 090501
  (2016).

\bibitem{fante2005electromagnetic}
R.~L. Fante, Electromagnetic beam propagation in turbulent media.
  \emph{Proceedings of the IEEE} \textbf{63}~(12), 1669--1692 (2005).

\bibitem{ghalaii2022quantum}
M.~Ghalaii, S.~Pirandola, Quantum communications in a moderate-to-strong
  turbulent space. \emph{Communications Physics} \textbf{5}~(1), 38 (2022).

\bibitem{trichili2020roadmap}
A.~Trichili, M.~A. Cox, B.~S. Ooi, M.-S. Alouini, Roadmap to free space optics.
  \emph{Journal of the optical society of America B} \textbf{37}~(11),
  A184--A201 (2020).

\end{thebibliography}
\bibliographystyle{sciencemag}

\section*{Acknowledgments}
We acknowledge support by the National Natural Science Foundation of China (No. 12205011 and No. 11974205),  the Young Elite Scientists Sponsorship Program by CAST (No. YESS20240786), Tsinghua University Initiative Scientific Research Program, and Beijing Advanced Innovation Center for Future Chip (ICFC).
\paragraph*{Author contributions:}
D.P. and G.-L.L. supervised and conceived the work. Z.-Z.S., Y.-B.C., Y.-C.L., D.P., and G.-L.L. designed and performed the experiments. Z.-Z.S., D.P, and Y.-B.C. analysed the experimental data and constructed the channel model. D.P. and J.G. designed the software. X.-T.S. and W.Z. contributed to the preliminary experiments. Z.-Z.S., D.P., and G.-L.L. wrote the paper with contributions from all authors. All authors discussed the results.
\paragraph*{Competing interests:}
The authors declare no competing interests.
\paragraph*{Data and materials availability:}
Correspondence and requests for materials should be addressed to Dong Pan and Gui-Lu Long.


\subsection*{Supplementary materials}
Secrecy capacity in cross-medium transmission\\
Figs. S1 to S2\\
Tables S1 to S2\\


\newpage


\renewcommand{\thefigure}{S\arabic{figure}}
\renewcommand{\thetable}{S\arabic{table}}
\renewcommand{\theequation}{S\arabic{equation}}
\renewcommand{\thepage}{S\arabic{page}}
\setcounter{figure}{0}
\setcounter{table}{0}
\setcounter{equation}{0}
\setcounter{page}{1} 


\begin{center}
\section*{Supplementary Materials for\\ \scititle}

	Ze-Zhou Sun$^{1,2,\ast}$, Yuan-Bin Cheng$^{1,2,\ast}$, Yu-Chen Liu$^{1,2,\ast}$, Jianxing Guo$^{1}$,\\ Xiao-Tian Song$^{1}$, Wei Zhang$^{1}$, Dong Pan$^{1,\dagger}$ and Gui-Lu long$^{1,2,3,4,\dagger}$\and
\and

	\small$^{1}$Beijing Academy of Quantum Information Sciences, Beijing 100193, China\and

	\small$^{2}$State Key Laboratory of Low-dimensional Quantum Physics and Department of Physics,\\ \small Tsinghua University, Beijing 100084, China\and

		\small$^{3}$Frontier Science Center for Quantum Information, Beijing 100084, China\and

			\small$^{4}$Beijing National Research Center for Information Science and Technology, Beijing 100084, China\and

	\small$^\dagger$Corresponding authors' email: pandong@baqis.ac.cn; gllong@mail.tsinghua.edu.cn\and

	\small$^\ast$These authors contributed equally to this work
\end{center}

\subsubsection*{This PDF file includes:}
Secrecy capacity in cross-medium transmission\\
Figs. S1 to S2\\
Tables S1 to S2\\
\newpage


\subsection*{Secrecy capacity in cross-medium transmission}

\label{Suppl:A}
Gaussian beams are employed to characterize the spatial profile of weak coherent pulse sources, which sufficiently describe the link loss in free-space optical transmission and have been widely adopted in the Bennett-Brassard 1984 and entanglement distribution protocols~\cite{karakosta2025free,bourgoin2015free}. We employ the widely adopted metric, the Rytov number $\sigma_{Ry}^2$, to quantify the turbulence intensity within a link~\cite{vasylyev2016atmospheric}. It characterizes the fluctuations in the received irradiance or in the phase and amplitude of the light as it propagates through turbulent media, with its definition given by 
\begin{eqnarray}
  \begin{split}
  \label{eq:1}
  \sigma _{Ry}^2=1.23C_n^2k^{\frac{7}{6}}d_1^{\frac{11}{6}}  ,
\end{split}
\end{eqnarray}
where $k= 2\pi/\lambda$ is the wavenumber, $d_1$ is the distance of free-space channel, and $C_n^2$ is the atmospheric index-of-refraction structure constant, which is used to quantify the refractive index variations caused by inhomogeneities in temperature, humidity, pressure, and other factors in the atmosphere. One parameter associated with photon propagation in turbulent channel is~\cite{fante2005electromagnetic}
$d_i = 1/(C_n^2 k^2 l_0^{5/3})$.
It denotes the propagation length at which the transverse coherence radius of a light wave becomes comparable to the turbulence inner scale $l_0$, which characterizes the smallest length scale over which refractive-index fluctuations remain correlated. As shown by the parameters in Table~\ref{Tab:1}, $d_i$ is approximately 475 km under our experimental conditions.

Assuming that the beam waist of Alice’s weak coherent pulse is $w_0$, the telescope magnification is $\Gamma$, and the light is collected by a receiver with a finite aperture of radius $a_R$. It has been proven that~\cite{ghalaii2022quantum}, in both weak and moderate-to-strong turbulence, provided the transmission distance $d_1$ is less than the critical propagation distance of $d_i$, the effective beam waist $w_{d_1}$ becomes
\begin{eqnarray}
  \begin{split}\label{eq:2}
  w_{d_1}=w_0 \Gamma \sqrt{1+(\frac{d_1}{d_R})^2}\sqrt{1+1.63(\sigma _{Ry}^2)^{\frac{6}{5}}\Theta },
\end{split}
\end{eqnarray}
where $d_R=\pi w_0^2\Gamma^2/ \lambda$ is the Rayleigh length of the beam, and $\Theta$ is defined as
\begin{eqnarray}
  \begin{split}\label{eq:3}
  \Theta=\frac{2d_1d_R^2}{kw_0^2\Gamma^2(d_R^2+d_1^2)}.
\end{split}
\end{eqnarray}
Then, the total transmittance $\eta$ for the entire link, which accounts for the free-space channel loss, conversion loss, fiber channel loss, the transmittance of receiving end $\eta_B$, and the detector efficiency $\eta_D$, can be expressed as~\cite{ghalaii2022quantum}
\begin{eqnarray}
  \begin{split}\label{eq:4}
  \eta=(1-e^{-\frac{2a_R^2}{w_{d_1}^2}})10^{-\frac{\alpha_1 d_1}{10}}10^{-\frac{L_c}{10}}10^{-\frac{\alpha_2 d_2}{10}}\eta_B \eta_D,
\end{split}
\end{eqnarray}
where $\alpha_1$ and $\alpha_2$ respectively denote the attenuation coefficients of the free-space channel and the fiber channel, $d_2$ is the distance of fiber channel, and $L_{c}$ is the conversion loss, which includes the loss associated with the transmitting telescope converting the optical signal in fiber into a spatial beam, and the receiving telescope coupling the spatial beam back into an optical fiber.

The secrecy capacity of the STIKE protocol based on the decoy-state method is given by~\cite{pan2025simultaneous}
\begin{eqnarray}
  \begin{split}\label{eq:6}
  C_s=q Q_\mu \left \{-f H_2(E_\mu)+\frac{Q_1}{Q_\mu}\left [ 1-H_2(e_1) \right ]   \right \} ,
\end{split}
\end{eqnarray}
where $H_2(x)=-x\log(x)-(1-x)\log(1-x)$ is the binary entropy function, $q=1/2$, and $f$ is the error correction efficiency. Furthermore, $Q_\mu$ and $E_\mu$ denote the total gain and QBER of the signal state, respectively, while $Q_1$ and $e_1$ represent the gain and QBER of the single-photon state. These quantities are expressed as follows:
\begin{eqnarray}
  \begin{split}\label{eq:7}
  Q_{\mu(\nu)}=Y_0+1-e^{-\eta \mu (\nu)},\\
  E_{\mu(\nu)}Q_{\mu(\nu)}=e_0 Y_0+e_{\rm det} (1-e^{-\eta \mu (\nu)}),\\
  Q_1=\frac{\mu^2e^{-\mu}}{\mu\nu-\nu^2}(Q_\nu e^\nu-Q_\mu e^\mu\frac{\nu^2}{\mu^2}-\frac{\mu^2-\nu^2}{\mu^2}Y_0),\\
  e_1=\frac{\mu e^{-\mu}(E_\nu Q_\nu e^{\nu}-e_0 Y_0)}{\nu Q_1},
\end{split}
\end{eqnarray}
where $\mu$, $\nu$ are the average photon numbers of the signal state and the decoy state, respectively. $Q_\nu, E_\nu$ are the total gain and QBER of the decoy state, and $Y_0 =2p_d$ is the yield of a zero-photon state. $e_{\rm det}=(1-V)/2$ is the intrinsic detector error rate, where $V$ is the interference visibility, and $e_0=1/2$ is the error rate of background. The quasi-QSDC and QKD in STIKE share the same information-theoretic security bound.

\section{Parameters obtained from the experiment}
\label{sec:B}
  
\begin{figure}[t]
\centerline{\includegraphics[width=\linewidth]{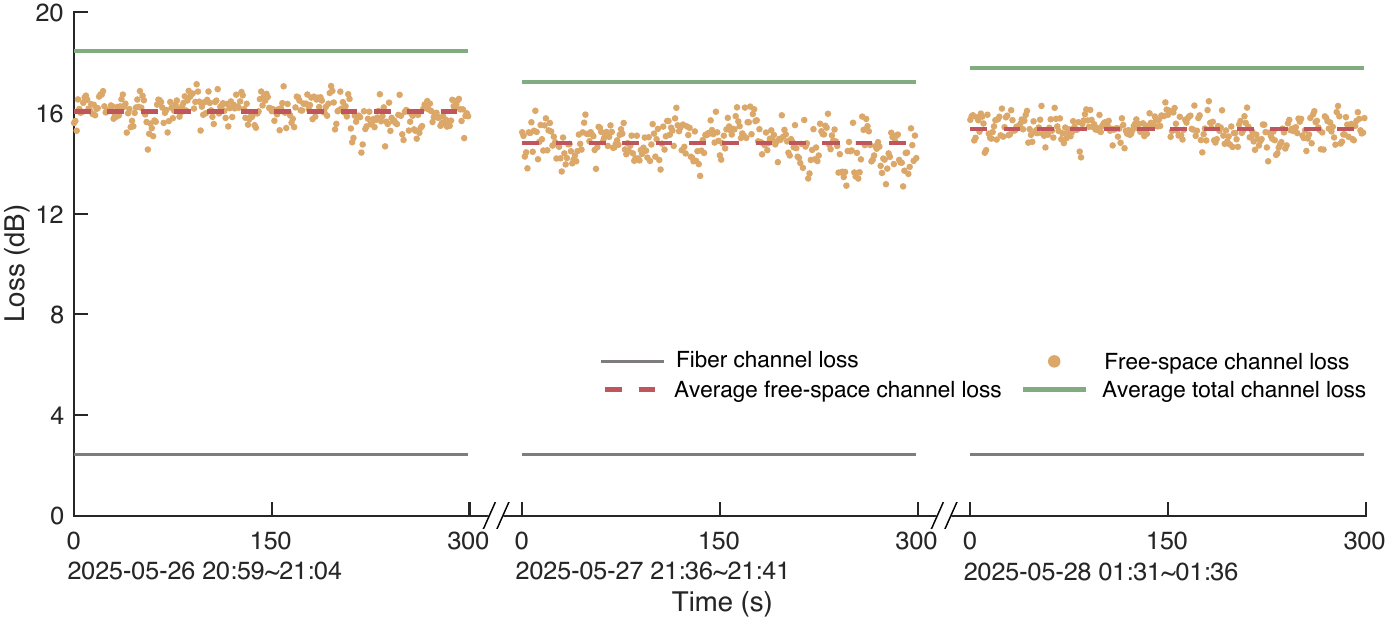}}
\caption{\label{Fig_loss} 
Temporal variation of total loss in the 1.4 km free-space and 10 km fiber hybrid channel. The channel loss of the 10-km fiber is 2 dB, and the loss of its associated fiber optic adapter is 0.42 dB. The total channel loss is the sum of the free-space channel loss and the fiber channel loss (2.42 dB).}
\end{figure}

\begin{figure}[t]
\centerline{\includegraphics[width=14cm]{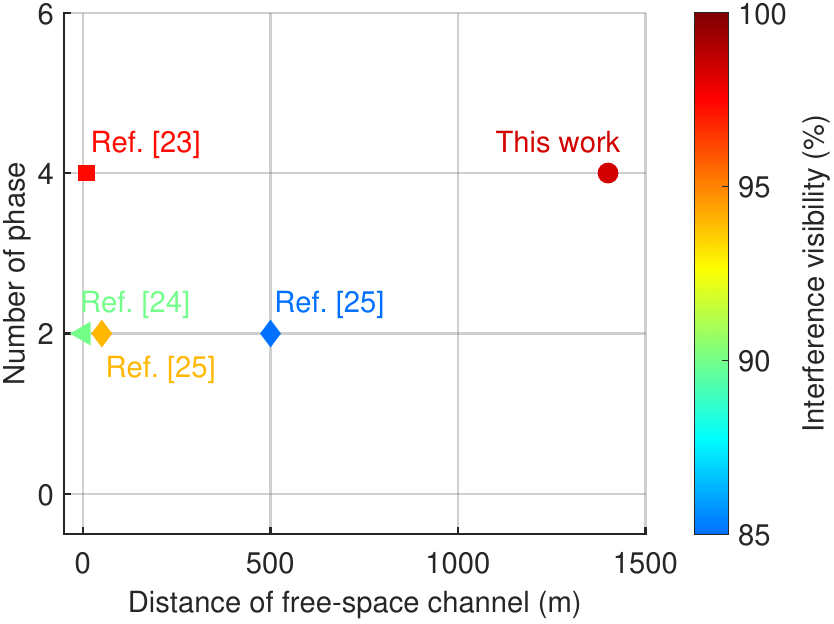}}
\caption{\label{Fig_INT} 
Interference visibility of our system compared with previous results~\cite{pan2020experimental,sajeed2021observing,cocchi2025time}, all without an APT system.}
\end{figure}

Extended Data Fig.~\ref{Fig_loss} shows the variation of the total channel loss $L_t$ over time for the 1.4 km free-space link combined with 10 km of optical fiber during three 5-minute intervals. The first segment of loss data was obtained at the start of the first experiment, while the second and third segments were measured before and after the second transmission experiment, respectively. The total loss from the 10-km optical fiber and the fiber optic adapter is 2.42 dB. The maximum loss fluctuations across the three intervals are 2.72 dB, 3.17 dB, and 2.39 dB, respectively. The average total channel loss over the two-day experiment was 18.48 dB and 17.51 dB, respectively. The relatively low average channel loss on the second day, together with the consistency of the loss measurements before and after the experiment, demonstrates the potential for our system to operate continuously for extended periods even without an APT system actively maintaining link quality. In free-space transmission of time-bin–encoded qubits, quantum states in the X basis intrinsically possess two distinct phases and are ultimately measured through interference. The system performance is therefore characterized by the interference visibility. We compare our results with two representative studies~\cite{sajeed2021observing,cocchi2025time}, even if they are not pure phase-encoded qubit transmission over free-space channel, as shown in Extended Data Fig.~\ref{Fig_INT}. Extended Data Tables 1 and 2 summarize the measured experimental parameters and results, as well as the simulation parameters.

\begin{table}[htbp]
\centering
\caption{Extended Data Table 1: Experimental parameters and results.}
\label{Tab:1}
\begin{tabular}{l c c}
\hline
\textbf{Description} & \textbf{Parameter} & \textbf{Value} \\
\hline
Telephoto system magnification & $\Gamma$ & 27.1$\times$ \\
Receiver entrance pupil radius & $a_R$ & 60 mm \\
Error correction efficiency & $f$ & 1.22 \\
Dark count rate of single-photon detector & $p_d$ & $1 \times 10^{-6}$ \\
Internal transmittance at Bob's side & $\eta_\mathrm{B}$ & $10^{-6.5/10}$ \\
Average photon number of signal state & $\mu$ & 0.71 \\
Average photon number of decoy state & $\nu$ & 0.28 \\
Detector efficiency & $\eta_D$ & 20\% \\
Two-day average interference visibility & $V$ & 98.47\% \\
Conversion loss &$L_{c}$& 15.40 dB \\
Total channel loss& $L_t$ & 17.83 dB \\
QBER of signal state & $E_\mu$ & 2.87\% \\
QBER of decoy state & $E_\nu$ & 4.01\% \\
Signal-state gain& $Q_\mu$ & $5.31\times10^{-4}$ \\
Decoy-state gain& $Q_\nu$ & $2.09\times10^{-4}$ \\
\hline
\end{tabular}
\end{table}

\begin{table}[htbp]
\centering
\caption{Extended Data Table 2: Simulation parameters.}
\label{Tab:2}
\begin{tabular}{l c c}
\hline
\textbf{Description} & \textbf{Parameter} & \textbf{Value} \\
\hline
Emission beam waist & $w_0$ & 1.74 mm \\
Index-of-refraction structure constant & $C_n^2$ & $1.28\times10^{-14}\;\rm{m}^{-\frac{2}{3}}$~\cite{ghalaii2022quantum} \\
Turbulence inner scale  & $l_0$ & 0.001 m~\cite{ghalaii2022quantum}\\
Attenuation coefficient of free-space channel & $\alpha_1$ & 0.2 dB/km~\cite{trichili2020roadmap} \\
 Attenuation coefficient of fiber channel  & $\alpha_2$ & 0.2 dB/km  \\
\hline
\end{tabular}
\end{table}

\end{document}